\documentstyle[preprint,aps,floats,epsf]{revtex}

\def\ll{\ell\bar\nu \bar\ell \nu}
\def\etmiss{{\overlay{/}{E}}_T}

\newcommand{\notE}{\ \hbox{{$E$}\kern-.60em\hbox{/}}}
\newcommand{\notp}{\ \hbox{{$p$}\kern-.43em\hbox{/}}}
\def\D0{\mbox{D\O}}

\def\NPB#1#2#3{Nucl. Phys. {\bf B#1}, #3 (19#2)}

\def\PRD#1#2#3{Phys. Rev. D {\bf #1}, #3 (19#2)}
\def\PRL#1#2#3{Phys. Rev. Lett. {\bf #1}, #3 (19#2)}

\begin{document}


\baselineskip=18pt

\preprint{
\font\fortssbx=cmssbx10 scaled \magstep2
\hbox to \hsize{
\hfill$\vcenter{
\hbox{\bf MADPH-99-1140}
\hbox{\bf UCD-99-21}
\hbox{\bf hep-ph/9910500}
\hbox{October 1999}}$ }
}

\title{\vspace*{2cm}
Physics Potential of a Tevatron Tripler}
\author{V. Barger$^a$, K. Cheung$^b$, T. Han$^a$, 
C. Kao$^a$, T. Plehn$^a$, R.-J. Zhang$^a$}
\address{${}^a$Department of Physics, University of Wisconsin, 
1150 University Avenue, Madison, WI 53706}
\address{${}^b$Department of Physics, University of California, 
Davis, CA 95616}


\maketitle

\thispagestyle{empty}

\begin{abstract} 
We explore the capabilities for new physics discovery 
in proton-antiproton collisions at 5.4~TeV center-of-mass energy 
with luminosity $10^{33}\rm\, cm^{-2}\, s^{-1}$ 
at a Tripler upgrade of the Tevatron collider. 
The prospects are robust for the usual Higgs boson 
and supersymmetry benchmarks. 
With an integrated luminosity of 40 fb$^{-1}$, 
discoveries at 5$\sigma$ could be made for 
a standard Higgs boson of mass $\alt 680$~GeV 
(600 GeV for 10 fb$^{-1}$), 
a lighter chargino of mass $\alt 380$~GeV, 
and an extra $Z$ boson of mass $\alt 2.6$~TeV; 
four-fermion contact interactions from new physics with scale 
$\alt 74$~TeV could be excluded at the 95\% 
confidence level. 
\end{abstract}

%


\section{Introduction}

Recent developments in superconducting 
magnet design may make it possible to upgrade the Fermilab Tevatron collider 
at moderate cost by replacing its ring of 4 Tesla dipole magnets 
by 12 Tesla dipoles\cite{tripler}. 
This Tripler design would yield proton-antiproton colliding beams 
at a center of mass energy $\sqrt s = 5.4$~TeV 
and a luminosity about $10^{33} \rm\, cm^{-2}\, s^{-1}$, 
which could significantly enhance the potential for new particle discoveries 
at the Tevatron. 
The physics justification for the Tripler requires detailed benchmark studies 
and comparison with the LHC potential, since the LHC is expected 
to become operational on a similar time-line as the Tripler. 
In this Letter we present benchmark results 
for discovery of the Higgs boson in the Standard Model (SM), 
the trilepton and same-sign dilepton signals 
of supersymmetric (SUSY) particles, extra $Z$-bosons 
and contact interactions. 
The interesting range for the SM Higgs boson mass is 
102.6 GeV $\alt m_H \alt$ 550 GeV;
the lower value is the current LEP 2 experimental limit \cite{LEP2}; 
the upper value is due to the requirement of a non-trivial 
and consistent effective theory\cite{Quigg}. 
Lattice calculations give a mass bound of $m_H \alt 710$ GeV \cite{Lattice}. 
Electroweak precision data favor $m_H \alt 255$ GeV at 95\% C.L. 
\cite{Precision}. 
The lightest Higgs scalar in minimal supersymmetry (SUSY) is predicted 
to have a mass less than 130 GeV \cite{Higgs} and our SM analysis 
applies equally to it in the decoupling limit \cite{Haber}.
Naturalness considerations in the minimal supergravity (SUGRA) model 
lead to an upper bound on the lighter chargino mass of 
$m_{\chi^\pm_1} \alt$ 400 GeV \cite{Naturalness}.

We show below that the Tevatron Tripler with ${\cal L} = 40$ fb$^{-1}$ 
would provide at least a 5$\sigma$ discovery for the SM Higgs boson 
up to $m_H = 680$ GeV and for the lighter chargino 
up to $m_{\chi^\pm_1} = 380$ GeV.

\section{Higgs bosons} 

\begin{figure}[htb]
\centering\leavevmode
\epsfxsize=4.0in\epsffile{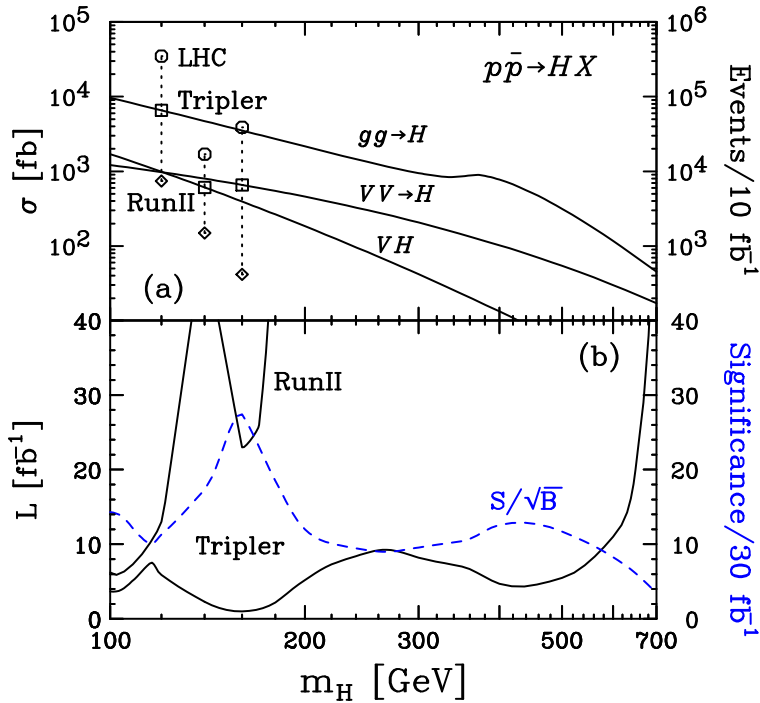}
\smallskip
\caption[]{
(a) The Higgs boson production cross section 
via gluon fusion, vector-boson fusion and associated production processes
versus $m^{}_H$ at a 5.4~TeV Tripler. 
Also shown are representative production cross sections 
at the Tevatron Run II (diamond), the Tripler (square) and the LHC (circle).
(b) The required integrated luminosity to reach $5\sigma$ statistical
significance versus $m^{}_H$ (solid curves). 
The lower and upper curves correspond to the Tripler 
and the Tevatron Run II respectively.
(The Tevatron curve is taken from 
the Run II Higgs Boson Working Group report \cite{HWG}).
Five channels (see text for details) have been combined for the 
Tripler discovery curve. The dashed curve gives
the signal significance $S/\sqrt B$
at the Tripler for 30 fb$^{-1}$ (the right-hand scale).
}
\label{fig:higgs}
\end{figure}

The Tevatron Tripler has great potential 
for the Standard Model Higgs boson search. 
In Fig.~\ref{fig:higgs}(a) we show the fusion cross section 
from $gg\rightarrow H$ as well as $VV\rightarrow H$, 
and the associated production $p{\bar p}\rightarrow VH +X$ 
at $\sqrt s = 5.4$~TeV, where $V=W^\pm, Z$.
For comparison, we also show the cross sections at the Tevatron Run II 
($\sqrt{s} =$ 2~TeV) 
and the LHC for some representative Higgs boson masses. 
The scale on the right-hand side gives expected 
number of events for $10$ fb$^{-1}$ integrated luminosity. 
We see that the total cross sections for Higgs production are increased
by about an order of magnitude at the Tripler compared to 
the Tevatron Run II. As a remark, if the Tripler is running at
the $pp$ mode, the $gg\to H$ cross section would not change;
that for $VV\to H$ would be reduced by about 10\%
and that for $q\bar q\to VH$ by about 25\%.


For a Higgs boson mass $m^{}_H < 130$ GeV, it is known
that the leading signal at the Tevatron is 
$p\bar p \to W^\pm H \to \ell^\pm b\bar b X$ \cite{Scott}. 
However, this channel becomes less accessible at the LHC because of the
much larger QCD background. We studied this process at the Tripler
and found that it is a useful channel to reach 5$\sigma$ significance
for $m^{}_H\alt 120$ GeV with a luminosity of less than 8 fb$^{-1}$.

For the intermediate range mass Higgs boson ($125\alt m^{}_H\alt 220$ GeV),
the most important discovery channel is the signal of dileptons along with 
missing transeverse energy ($\etmiss$) through the process
$p{\bar p}\rightarrow gg\rightarrow H\rightarrow WW^*\rightarrow\ll$ 
($\ell = e$ or $\mu$).
The leading background is $WW$ production. 
We impose kinematic cuts similar to those in \cite{HZ} 
and find that this mode has a 5$\sigma$ statistical significance 
reach for $120\alt m^{}_H\alt 220$ GeV 
(assuming ${\cal L} =$ 20 fb$^{-1}$).
In comparison, at the Tevatron Run II with ${\cal L} = 30$~fb$^{-1}$, 
a $5\sigma$ significance can be reached 
only at the narrow mass range near $m^{}_H\simeq 160$~GeV \cite{HZ}. 

When $m^{}_H > 2 m^{}_Z$, the ``gold-plated'' channel
$H\rightarrow ZZ\rightarrow 4\ell$ becomes interesting. 
After the basic acceptance cuts and 
a search for a peak in the four-lepton invariant mass distribution, 
this channel has a good signal to background ratio, $S/B \sim 1$; 
for example, with $m_H = 220$ GeV, 
the signal and background cross sections are 1.2 fb and 1.5 fb, respectively.

For a still larger Higgs boson mass, the ``silver-plated'' channel
$H\rightarrow ZZ\rightarrow\ell\bar\ell\nu\bar\nu$ is more useful 
since its rate is 6 times larger than the $4\ell$ rate, 
although Higgs boson mass reconstruction is more difficult 
because of the missing neutrinos. 
The irreducible background is $ZZ$ production. 
After suitable cuts on $\notE_T$ and 
the $M_C(\ell\ell,\etmiss)$ cluster transverse mass, 
we find the 5$\sigma$ discovery reach for $m_H$ 
is about 500 GeV for ${\cal L}=$ 20 fb$^{-1}$.

For a Higgs boson mass $m_H \agt 500$ GeV, the $H \to WW \to \ell\nu jj$ 
channel can also be used. 
The leading background is the $Wjj$ QCD process.
We find that 5$\sigma$ discovery reach of this channel 
can be achieved for $m_H \approx 650$ GeV for ${\cal L} = 20$ fb$^{-1}$.

In Fig.~\ref{fig:higgs}(b), we plot the $5\sigma$ mass reach for Higgs
boson discovery at the Tripler from combining the above channels
(the lower solid curve). With ${\cal L} = 20$ fb$^{-1}$, 
a 5$\sigma$ Higgs boson discovery would be possible over the 
entire mass range from the current limit
of 102.6 GeV all the way up to 650 GeV. In comparison, 
with ${\cal L} = 30$ fb$^{-1}$, the Tevatron Run II 
(the upper solid curve) can only cover a very limited range 
of the Higgs boson masses, {\it i.e.}, $m^{}_H < 130$ 
GeV and $m^{}_H\simeq 160$ GeV. The dashed curve
gives the signal significance $S/\sqrt B$
at the Tripler for 30 fb$^{-1}$.
It is important to see that for $m^{}_H < 200$ GeV we should be
able to achieve more than 10$\sigma$ significance via 
$WH(\to b\bar b)$ and $gg\to H\to WW$, both involving 
the electroweak gauge coupling $WWH$.
We note that further improvement of the reach at the Tripler is possible 
by optimizing acceptance cuts and by including the $ZH$ channel as well as 
channels involving tau-leptons in the final states. 
To summarize, the Tevatron Tripler offers a tremendous opportunity 
to discover the SM Higgs boson over the mass range 
from the current bound to $\sim 650~(680)$ GeV, 
for ${\cal L}=$ 20 (40) fb$^{-1}$.

\section{Supersymmetric particles}

A large number of production 
processes for supersymmetric particles have been experimentally investigated 
at the Tevatron, yielding lower mass limits for strongly interacting gluinos,
 
squarks, especially stops, and weakly interacting sleptons and
neutralinos/charginos \cite{SUGRA}. 
The signals of supersymmetric particles are trileptons 
from chargino/neutralino pairs, same-sign dileptons 
from gluino pairs and missing transverse energy carried by 
the lightest neutralino.

\begin{figure}[htb] 
\centering\leavevmode
\epsfxsize=4.5in\epsffile{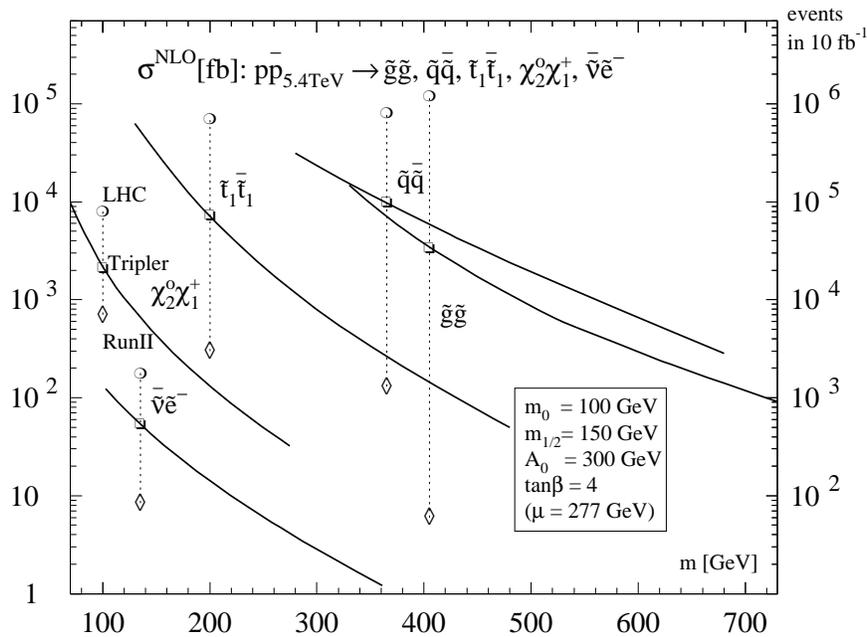}

\smallskip
\caption[]{\label{fg:prospino} 
Some next-to-leading order pair production cross sections 
versus the (smaller of the two) final state masses. The 
vertical bars indicate the SUGRA inspired scenario and show the
respective cross sections at the Tevatron Run II and the LHC.}
\end{figure}

Figure~\ref{fg:prospino} shows that the Tripler cross sections 
are considerably larger than those for the Tevatron Run II 
\cite{prospino}. 
The gluon fusion contribution to SUSY particle production 
at $\sqrt{s} =$ 5.4 GeV 
is more important than that of quark-antiquark annihilation.
The fractions of SUSY particle production from incoming gluons 
at the given SUGRA point are $67\%$ for stops and $72\%$ for gluino pairs 
at the Tripler, while they are only $25\%$ and $10\%$ for Run II.
This affects the $K$ factors due to SUSY-QCD corrections 
through the larger color charge of the radiating incoming partons. 


To assess the discovery potential of the Tripler 
in searching for SUSY particles, we present results from simulations 
for the trilepton and the same sign dilepton (SS2L) signals
with an event generator and a simple calorimeter including our acceptance cuts.
The ISAJET 7.40 event generator program \cite{ISAJET}
with the parton distribution functions of CTEQ3L \cite{CTEQ3L}
is employed to calculate the trilepton ($3\ell +\notE_T$) and the same sign 
dilepton (SS2L) signals from all possible sources of SUSY particles.
The backgrounds from $t\bar{t}$ is calculated with ISAJET as well.

The trilepton signature with missing transverse energy ($3\ell+\notE_T$) 
is one of the most promising 
channels to search for supersymmetric particles \cite{Trilepton1}.
At the Tevatron with $\sqrt{s} =$ 2 TeV, 
the major source of trileptons is the associated production 
and decay of the lighter chargino ($\chi^\pm_1$) 
and the second lightest neutralino ($\chi^0_2$) \cite{Trilepton2,Trilepton3}.
At the Tripler, 
gluinos and squarks can be the dominant source 
of trileptons and same-sign dileptons along with jets in the final states, 
while $\chi^\pm_1 \chi^0_2$, and the slepton pair production processes 
$\tilde{\ell}^*\tilde{\ell}$, 
and $\tilde{\ell}\tilde{\nu}$ generate clean trilepton events.

Requiring $p_T(\ell_{1,2,3}) > 20, 15, 10$ GeV, 
$|\eta(\ell_{1,2,3})| < 1,2,2$, 
and applying other acceptance cuts \cite{Trilepton2}, 
we find that the major SM backgrounds are 
 (i) $q\bar{q}' \to WZ +W\gamma \to \ell\nu \ell\bar{\ell}$
or $\ell'\nu' \ell\bar{\ell}$ ($\ell = e$ or $\mu$)
(ii) $q\bar{q}' \to WZ +W\gamma \to \ell\nu \tau\bar{\tau}$
or $\tau\nu \ell\bar{\ell}$ and subsequent $\tau$ leptonic decays, 
with one or both gauge bosons being virtual. 
We employed the programs MADGRAPH \cite{Madgraph}
and HELAS \cite{Helas} to evaluate
the background cross section of $p\bar{p} \to 3\ell +\notE_T +X$
for contributions from all these subprocesses.
Additional backgrounds come from production of $t\bar{t}$ 
and $ZZ \to \ell\bar{\ell}\tau\bar{\tau}$ \cite{Trilepton2,Trilepton3}.
At the Tripler with 20 fb$^{-1}$ integrated luminosity, 
we expect about 36 background events; 
a $5 \sigma$ signal would have 30 events. 
If the Tripler runs with $pp$ collisions, 
the major trilepton background cross section will be reduced by about 35\%
and the SUSY trilepton signal cross section 
will be reduced by about 18\%, 27\%, 55\% and 59\% 
for $m_{1/2} =$ 140, 200, 300 and 460 GeV, respectively. 

\begin{figure}[htb]
\centering\leavevmode
\epsfxsize=3.4in\epsffile{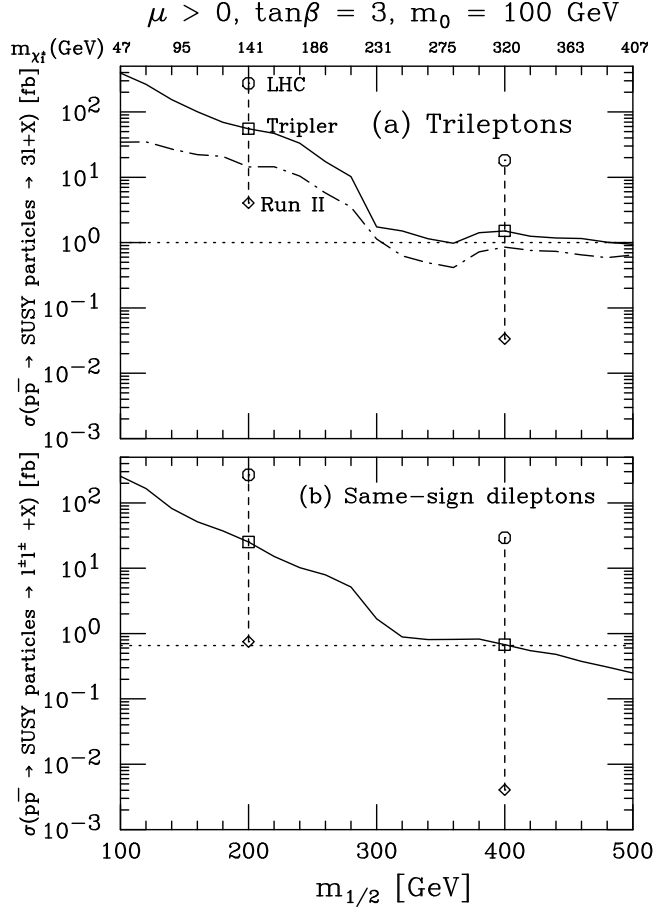}

\caption[]{
Cross sections of $p\bar{p} \to {\rm SUSY \; particles} \to 3\ell +X$
and $p\bar{p} \to {\rm SUSY \; particles} \to \ell^\pm\ell^\pm +X$
after cuts, versus $m_{1/2}$, at $\sqrt{s} = 5.4$ TeV,
with $\mu > 0$, $\tan\beta =$ 3, $m_0 =$ 100 GeV,
for (a) inclusive trileptons (solid) and clean trileptons (dash-dot), and 
(b) for same-sign dileptons with at least 2 jets (solid). 
Also shown are the cross sections at the Tevatron Run II (diamond) 
and the LHC (circle).
The dotted horizontal lines denote the 5$\sigma$ signal cross sections 
for ${\cal L} =$ 40 fb$^{-1}$. 
\label{fig:susy}}
\end{figure}

Figure \ref{fig:susy}(a) presents the cross sections at the Tripler
for trileptons with jets and for {\it clean} trileptons with 0 or 1 jet 
at the Tripler versus $m_{1/2}$, 
for $\tan\beta = 3$, $m_0 = 100$ GeV, and $\mu > 0$. 
Also shown is the cross section for a 5$\sigma$ signal 
with ${\cal L} = 40$ fb$^{-1}$.
The trilepton signal will be observable up to $m_{1/2} = 470$ GeV 
($m_{\chi^\pm_1} \sim$ 380 GeV, $m_{\tilde{g}} \sim$ 1.1 TeV) 
with ${\cal L} = 40$ fb$^{-1}$, 
which is a great improvement from the Tevatron Run II reach of 
$m_{1/2} \alt$ 260 GeV 
($m_{\chi^\pm_1} \sim 195$ GeV, $m_{\tilde{g}} \sim$ 600 GeV) 
with ${\cal L} = 30$ fb$^{-1}$ \cite{Trilepton2}.


Same-sign dileptons with missing energy (SS2L) 
is another promising channel to search for supersymmetry
\cite{SS2L,LHC,Tev2}.  
The cross section of the SM background in this channel is very small 
after suitable cuts.
There are several major sources of the SS2L signal:
(i) $\tilde{g}\tilde{g}$ production, 
where $\tilde{g} \to \chi^\pm_1 +q\bar{q}'$ 
and $\tilde{g}\tilde{g} \to \ell^\pm \ell^\pm +{\rm jets}\;\; +\notE_T$; 
these can test the Majorana property of the gluinos 
if $m_{\tilde{g}} < m_{\tilde{q}}$. 
(ii) $\tilde{g}\tilde{q}$; (iii) $\tilde{q}\chi^\pm$; 
(iv) $\bar{q}\bar{q}$ where $\tilde{q} \to q' \chi^\pm_1$; and
(v) trilepton events where one of the leptons is lost.

For same-sign dilepton events having $p_T > 15$ GeV 
[$p_T > 10$ (20) GeV for the Tevatron (LHC)] 
and $|\eta_{1,2}| < 1, 2$, we require at least two jets as well as 
the same isolation and missing energy as in the trilepton analysis.
The major SM backgrounds are 
(i) $WZ \to \ell\nu \tau\bar{\tau}$, (ii) $t\bar{t}$ and (iii) $ZZ$. 
The contributions from $WW$, $W+{\rm jets}$ and $Z+{\rm jets}$ 
are negligible after cuts \cite{LHC,Tev2}.
With ${\cal L} = 20$ fb$^{-1}$ at the Tripler, 
we expect about 14 background events; 
a $5 \sigma$ signal would have 19 events.
In $pp$ collisions at $\sqrt{s} =$ 5.4 GeV, 
the major same-sign dilepton background cross section from 
$WZ \to \ell\nu \tau\bar{\tau}$ will be reduced by about 30\%
and the SUSY same-sign dilepton signal cross section 
will be about the same as that in $p\bar{p}$ collisions. 

In Fig. 3(b), we present the cross sections after cuts 
versus $m_{1/2}$, for same-sign dileptons 
with $\tan\beta = 3$, $m_0 = 100$ GeV, and $\mu > 0$.
This signal is observable up to $m_{1/2} = 300$ GeV 
($m_{\chi^\pm_1} = 250$ GeV, $m_{\tilde{g}} = 780$ GeV) 
with ${\cal L} = 20$ fb$^{-1}$, 
and can be improved up to $m_{1/2} = 400$ GeV
($m_{\chi^\pm_1} = 320$ GeV, $m_{\tilde{g}} = 960$ GeV)
with ${\cal L} = 40$ fb$^{-1}$.
Other SUSY signals from single lepton+$\notE_T$ and 
$\notE_T$+jets, which are promising at the LHC \cite{LHC}, 
may also be useful at the Tripler.

\section{Contact interactions and extra $Z$-bosons}

Many sources 
of new physics at high scales can be parameterized by four-fermion 
contact interactions \cite{Contact1,Contact2},
\begin{equation}
L_{\rm NC} = \sum_q \sum_{\alpha,\beta = \rm L,R} 
{4\pi\epsilon\over (\Lambda_\epsilon^{\alpha,\beta})^2} 
\left(\overline{e_\alpha}\ \gamma_\mu e_\alpha\right) 
\left(\overline{q_\beta}\ \gamma^\mu q_\beta\right)
\end{equation}
where $\epsilon = \pm 1$ and $\Lambda$ is the typical energy scale at which 
the new physics sets in. 
The Drell-Yan process of lepton pair production at hadron colliders 
is a powerful tool to probe the contact scale of new physics contributions 
\cite{Contact2}. 
In this process the contact terms interfere with $\gamma$ and $Z$ exchanges, 
yielding multi-TeV sensitivities to contact scales $\Lambda$. 
The 95\% C.L. lower limits on $VV$ contact scales that can be set are
\begin{eqnarray}
&&\mbox{Tevatron \quad (2 TeV, 20 fb$^{-1}$) 
                 \qquad $\Lambda_+ = 36$ TeV} \nonumber\\
&&\mbox{Tripler  \qquad (5.4 TeV, 20 fb$^{-1}$) 
                 \quad $\Lambda_+ = 61$ TeV }
                 \label{eq:contact} \\  
&&\mbox{Tripler  \qquad ($pp$: 5.4 TeV, 20 fb$^{-1}$) 
                 \quad $\Lambda_+ = 34$ TeV } \nonumber\\
&&\mbox{LHC      \qquad \quad (14 TeV, 100 fb$^{-1}$) 
                 \quad $\Lambda_+ = 84$ TeV } \nonumber 
\end{eqnarray}
The comparative reach for other contact terms is similar.
With ${\cal L} = 40$ fb$^{-1}$ at the Tripler, 
contact interactions from new physics with a mass scale $\alt 74$~TeV 
could be excluded at 95\% C.L.

Additional $Z$-bosons are a generic feature of 
gauge symmetries larger than the SM. 
We consider two interesting models with extra $Z$ bosons: 
(i) $Z'_\chi$ in $\rm SO(10)\to SU(5)\times U(1)_\chi$, and 
(ii) $Z'_S$ in the sequential standard model, with SM coupling strength 
to quarks and leptons while couplings to $W$ and $Z$ bosons are suppressed.
The interactions of the extra $Z$ bosons with fermions 
are given by \cite{Zprime}
\begin{eqnarray}
{\cal L} = -g_2 \sum_i \bar{\psi}_i \gamma^\mu
(\epsilon_{iL} P_L +\epsilon_{iR} P_R) \psi_i Z_\mu',
\end{eqnarray}
where $P_{L,R} = (1\mp \gamma_5)/2$, 
$g_2 = \sqrt{5/3}\sin\theta_W g_1$,
and $g_1 = e/(\sin\theta_W \cos\theta_W)$. 
In the sequential standard model, 
the coupling constant in Eq. (3) is replaced by $g_1$ 
and $\epsilon_{L,R}$ are the same couplings 
as those of the SM $Z$ boson. We find that the 
5$\sigma$ reach limits are
\begin{eqnarray}
&&\mbox{Tevatron\quad  $M_{Z'_\chi} = 1.1$ TeV, $M_{Z'_S} = 1.2$ TeV }
                       \nonumber \\
&&\mbox{Tripler\qquad  $M_{Z'_\chi} = 2.6$ TeV, $M_{Z'_S} = 2.9$ TeV } \\
&&\mbox{Tripler\ ($pp$)\ \ $M_{Z'_\chi} = 2.0$ TeV, $M_{Z'_S} = 2.1$ TeV }
\nonumber \\
&&\mbox{LHC\qquad\quad $M_{Z'_\chi} = 5.1$ TeV, $M_{Z'_S} = 5.3$ TeV.}
                       \nonumber
\end{eqnarray}
The energies and luminosities of the colliders are the same 
as those in Eq.~(\ref{eq:contact}). 
Thus, the Tripler would more than double the $Z'_\chi$ reach of the Tevatron 
and have about half the reach of the LHC. 

\section{Conclusions} 

Our analyses of physics benchmarks show that 
the Tripler offers robust opportunities for the exploration 
of new physics. This is dramatically the case for the SM
Higgs sector where we find that with 40~fb$^{-1}$, a $5\sigma$ 
discovery can be made up to a Higgs mass of 680~GeV, 
thus covering the theoretical interesting mass range for the SM Higgs boson. 
In particular, a luminosity of 30~fb$^{-1}$ would lead to more
than 10$\sigma$ signal significance for $m_H^{}<200$ GeV via the
gauge coupling channels $WH$ and $gg\to H\to WW$, directly probing
the coupling of the gauge and Higgs bosons.
Supersymmetry searches at the Tripler via trileptons 
would extend up to a chargino mass $m_{\chi_1^\pm} = 380$~GeV, 
which is close to the naturalness upper bound. 
The scale of contact interactions can be pushed by more than 20~TeV 
above the expected reach of the Tevatron and extra $Z$ bosons could be 
discovered up to a mass 2.6~TeV.

Our comparisons of the Tripler and LHC shows that 
although the LHC has a greater new physics discovery reach in most channels, 
the Tripler can cover the interesting mass range for the SM Higgs 
and also offers coverage for most of the lighter chargino mass range 
expected from naturalness considerations. 
Thus, for the Higgs and chargino benchmarks both the Tripler and LHC 
may offer comparable physics opportunities. 

\section*{Acknowledgments} 

We would like to thank Peter McIntyre 
for inspiring us to undertake this study and 
Teruki Kamon for beneficial discussions. 
This research was supported in part by the U.S. Department of Energy 
under Grants No. DE-FG02-95ER40896 and No. DE-FG03-91ER40674, 
in part by the University of Wisconsin Research Committee
with funds granted by the Wisconsin Alumni Research Foundation 
and the Davis Institute for High Energy Physics.

\end{document}